# Marker enzyme phenotype ratios in agamospermous sugarbeet progenies as a demonstration of multidimensional encoding of inherited information in plants

E.V. Levites

Institute of Cytology and Genetics, Siberian Branch of the Russian Academy of Sciences, 10 Lavrentieva Ave. 630090 Novosibirsk, Russia e-mail: levites@bionet.nsc.ru

**It has been demonstrated that the observed ratio of phenotypes of marker enzymes in some sugarbeet plants produced by mitotic agamospermy can be explained by different degrees of endoreduplication of chromosomes carrying different alleles of the enzyme loci. In these plants, different patterns of variability of the enzymes controlled by the linked loci suggest different degrees of endoreduplication of different chromosomal regions. A concept of multidimensional encoding of inherited information in eukaryotes has been proposed.**
___________________________________________________________________________

## INTRODUCTION

Agamospermous reproduction in which seeds set without a pollen parent is widely spread in plants (Richards, 1986). The terms apomixis (Gustafsson, 1946, 1947; Asker and Jerling, 1992; Koltunow, 1993; Dresselhaus, 2000) and apozygoty (Khokhlov, 1967) are also used to designate the process. Under such reproduction, the embryo develops out of the cell which can be named apozygote (Maletskii et al., 2004). There are classifications for ways of reproduction considering the cytoembryological origin of cell entering embryogenesis (Gustafsson, 1946, 1947). The genetic classification considering the cell genome transformation before entering embryogenesis was suggested (Levites, 2002a). In this classification, agamospermy is considered as a subsexual process in which meiotic and mitotic agamospermy are distinguished. If the embryo develops out of the embryosac cell having formed after meiotic divisions of mother megaspore cell, then this process is called meiotic agamospermy previously designated as meiotic diplospory. Mitotic diplospory, apospory and adventitious embryogenesis are proposed to be referred to mitotic agamospermy. Herewith, embryo development proceeds out of the cell which has not undergone meiosis.

There is a wealth of evidence for variability in plants produced by mitotic agamospermy; however, theoretically, those progenies should be homogenous. The polymorphism revealed in these plants with marker enzymes was first associated with inactivation of either allele of the heterozygous enzyme loci (Levites et al., 1999a). However, it was later demonstrated that this polymorphism is more likely associated with another process which was preliminarily designated as redetermination of an enzyme locus (Levites, 2002b). Redetermination was conceived as failure by either of the two alleles of the maternal plant heterozygous locus to exhibit any activity in some part of agamospermously produced progeny; moreover, this part of progeny is homozygous on the allele that remains active.

Based on observations of the thing that colchicine treatment of maternal plants from which progeny was produced by agamospermy affects the ratio of phenotypic classes, it was hypothesised that chromosome endoreduplication is an important factor driving studied plant somatic cells into embryogenesis (Levites et al., 2000). This hypothesis was supported by the existence of diploid sugarbeet plants with high content of DNA in cell nucleus (Maletskaya and Maletskaya, 1999) and diploid plants of *Allium tuberosum* with endoreduplicational meiosis (Kojima and Nagato, 1992) capable of agamospermous reproduction. It is also proved by close association between agamospermy and polyploidy (Asker and Jerling, 1992) and simultaneous occurrence of both polyploid and diploid cells with endoreduplicated chromosomes in plant generative organs (Carvalheira, 2000).

Simultaneously, different variability observed in agamospermously produced plants at different alleles of the enzyme loci *Adh1* и *Idh3* located 17 cM apart on the same chromosome (Levites et al., 1988, 1994) suggests that the degree of endoreduplication is varying along the chromosome, i.e. the same chromosome can have highly and weakly polytenized regions (Levites, 2003, 2005).

The aim of this research was to obtain and to consider new experimental evidence in support of the earlier hypotheses and to develop a model with which the features of the structural and functional organization of the plant cell nucleus that determine the cell's genomic variability could be characterized.

## MATERIALS AND METHODS

Agamospermous progenies used in this research were developed on the basis of the method of growing blooming pollen-sterile plants in pollenless regime (Maletskii and Maletskaya, 1996). Pollination-free settings were made by cultivating pollen-sterile sugarbeet plants (ms1-5, ms11-3, msKWS1-5, ms2-5, ms12-11, ms SOAN-41 x red table beet) on a remote plot or under cotton bags preventing any pollen ingression. One of the progeny (12-3) had been produced by emasculation of a semifertile plant two or three days before bloom followed by insulation of shoots with emasculated buds. Each sample assayed was represented by the progeny of a single sterile plant.

Alcohol dehydrogenase (ADH1, E.C.1.1.1.1.), isocitrate dehydrogenase (IDH3, E.C.1.1.1.42.) and malic enzyme (ME1, E.C. 1.1.11.40.) controlled by loci *Adh1, Idh3 and Me1*, respectively, were used as marker characters for estimation of variability (Levites, 1979; Maletskii and Konovalov, 1985; Levites et al, 1994). Determining the genotype of an agamospermous progeny on enzyme genes was carried out either directly at the seed stage or later, - namely at the blooming stage in plants grown from agamospermous seeds. In the last case, determination of genotypes of analyzed plants was made by pollination with pollen of tester plants. Thus, genotypes of plants grown from agamospermous seeds were analysed on their progeny.

Seeds from each progeny were analyzed individually by starch gel horizontal electrophoresis using gel buffer 0.012M Tris-citrate, pH 7.0, and tray buffer 0.037M Tris-citrate, pH 7.0 (Meizel and Markert, 1967). For malic enzyme extraction, individual seeds weighing 1.5-2 mg were ground in 10-12 μl of 0.1M Tris-HCl buffer (pH 8.3). A 2.5x9 mm piece of Whatman 3MM paper was soaked in the obtained homogenate and inserted into a slit in the starch gel. Electrophoresis was performed at a voltage gradient of 6-7 v/cm, at $4^0$C for 18 h. Histochemical visualization of isozymes has been made using 0.05 M Tris-HCl, pH 8.3 containing 0.1 mM phenazine methosulfate, 0.3 mM thiazolyl blue tetrazolium bromide. As cofactors, 0.1 mg/ml NAD for ADH1, or 0.1 mg/ml NADP and 0.01 M $MgCl_2$ for IDH3 and ME1 were added in this mixture. In staining mixtures, 0.5% ethanol for ADH1, 25 mM L-malic acid Na-salt for ME1 or 0.6 mM DL isocitric acid $Na_3$ salt for IDH3 were used as substrates.

# RESULTS AND DISCUSSION

Polymorphism on ADH1, IDH3 and ME1 in sugarbeet plants produced by agamospermy was revealed. The phenotypes of plants homozygous on *Adh1*, *Idh3* and *Me1* are revealed in seeds as single-band isozyme patterns with a faster (FF) or a slower (SS) electrophoretic mobility (Fig. 1).

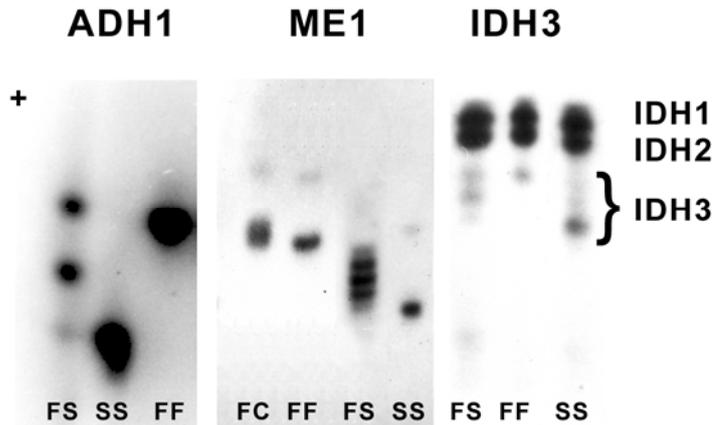

Fig.1 Isozyme patterns of alcohol dehydrogenase (ADH1), malic enzyme (ME1) and isocitrate dehydrogenase (IDH3) in sugarbeet seeds (Levites et al., 2001b). Migration is toward anode.

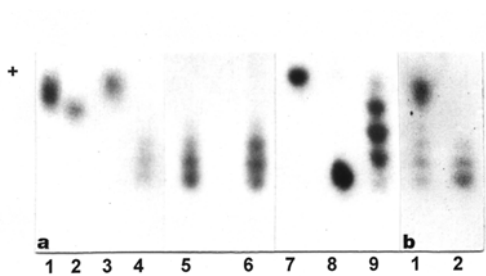

Fig.2. Isozyme patterns of malic enzyme in agamospermous sugarbeet seeds (Levites et al., 2001a). A: lanes 1 and 3, phenotype FC; lane 2, phenotype FF; lanes 4-6, phenotype FS; lane 7, phenotype CC; lane 8, phenotype SS; lane 9, phenotype CS. B – malic enzyme in KHBC2-3A seeds: lane 1, unusual phenotype (FC+FS); lane 2, phenotype FS. Migration is toward anode.

The heterozygous pattern (FC) for ME1 is revealed on electrophoregram as a wide enzyme activity zone if its five isozymes have weak electrophoretic distinctions (Fig. 1, 2). The heterozygous respect to the loci *Adh1* and *Idh3* plants feature a three-band electrophoretic pattern (FS) for ADH1 and IDH3. Significant differences in electrophoretic mobility of ADH1 isozymes are conditioned by differences in two aminoacids and in two electrons in enzyme allelic subunits, respectively (Vinichenko et al., 2004). The distinct character of ADH1 isozyme pattern makes it a most convenient model for studying the allele expression of isozyme locus. Isozyme band intensity ratios in heterozygous ADH1 patterns of studied agamospermous progenies were practically the same and similar with that of sexually produced diploid heterozygotes (Fig. 3). It is indicative of the thing that elementary single doses of isozyme locus alleles are expressed in heterozygotes of both gamo- and agamospermous progenies. In single cases amplification of this or that *Adh1* allele relative expression was observed.

A   B

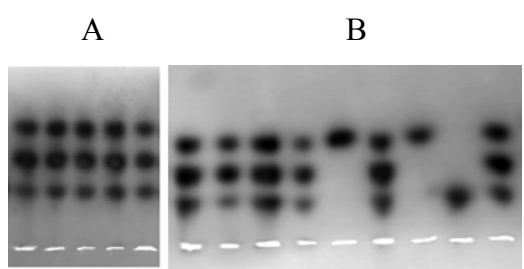

Figure 3. ADH isozyme patterns in gamospermously (A) and agamospermously (B) produced sugarbeet seeds.

The ratios of phenotypic classes of marker enzymes ADH1 and IDH3 are presented in Table 1. This table has no entries for seeds with null enzymic activity, which are frequently encountered in plants produced by agamospermy. Data on null phenotypes for ADH1 are presented in Table 3. In progenies ms1-5 and 12-3, marker enzymes ADH1 and IDH3 are presented by three phenotypic classes, whereas other progenies have two of them (Table 1).

This dimorphism is the most unique feature of progenies produced by mitotic agamospermy (Levites et al., 1999a). As it can be easily seen, the ratio of phenotypic classes for ADH1 and IDH3 within the same progeny is not the same (Table 1). Therefore, this points out the fact that variability on the *Adh1* locus and variability on the *Idh3* locus of the seeds produced by agamospermy are rather independent phenomena. Results of analysis of the digenic phenotypes for these enzymes (Table 2) provide further support to this conclusion; ADH1 phenotypes are designated FF, FS and SS, IDH3 phenotypes - ff, fs and ss.

Table 1. Marker enzyme phenotype ratios in sugarbeet seeds set by agamospermy

| Progeny | ADH1 FF-FS-SS | IDH3 FF-FS-SS | References |
|---|---|---|---|
| ms1-5 | 3-19-6 | 4-17-7 | This work |
| ms11-3 | 4 -4 -0 | 5-3-0 | -//- |
| MsKWS1-5A | 0-78-0 | 23-41-0 | Levites et al., 1999a |
| ms2-5 | 46-39-0 | 38-46-0 | Levites et al., 2001b |
| 12-3 (emasculated) | 8-6-3 | 7-6-3 | -//- |
| ms12-11 | 9-4-0 | 7-6-0 | -//- |

Table 2. Ratios of digenic phenotypes for ADH1(FF,FS,SS) and IDH3(ff,fs,ss) in sugarbeet seeds set by agamospermy

| Progeny | FFff | FFfs | FFss | FSff | FSfs | FSss | SSff | SSfs | SSss |
|---|---|---|---|---|---|---|---|---|---|
| ms1-5 | 2 | 1 | 0 | 2 | 15 | 2 | 0 | 1 | 5 |
| ms11-3 | 3 | 1 | 0 | 2 | 2 | 0 | 0 | 1 | 5 |
| MsKWS1-5A | 0 | 0 | 0 | 25 | 40 | 0 | 0 | 0 | 0 |
| ms2-5 | 24 | 22 | 0 | 14 | 23 | 0 | 0 | 0 | 0 |
| 12-3 (emasculated) | 5 | 2 | 1 | 2 | 3 | 1 | 0 | 1 | 2 |
| ms12-11 | 7 | 3 | 0 | 0 | 3 | 0 | 0 | 0 | 0 |

Curious, but under this type of agamospermy, plants develop from cells that have not undergone the genomic reorganisation typical of meiosis. Hence, there are two questions: what is it that underlies this polymorphism, and why are these two genetic and enzyme systems each on its own way of variability? Let us consider some previously published data. Table 3 presents a portion of data obtained earlier from the analysis of polymorphism for ADH1 sugar beet seeds produced by agamospermy (Levites et al., 2000).

Plants maternal to some of these progeny were treated with a low-concentrated colchicine solution at the stage of germination. Colchicine concentration was sufficient for some stimulation cell nuclear DNA production, yet insufficient to give rise to perfect tetraploids (Maletskii, 1999). Other progenies were used as controls (Table 3). The respective ratios of phenotypes FF and FS in experimental progeny msKW-88c and control progeny msKW-35 are 1:1. The respective ratios of phenotypes in control progenies msKW-11 and msKW-13 and experimental progeny msKW-11c are also close to one another. Therefore, there is some logic behind this polymorphism in plants produced by agamospermy.

Table 3. Ratios of phenotypes on the *Adh1* locus in agamospermous progeny in control and colchicine treated ($C_0$) sugarbeet hybrid plants ms Klein Wanzleben (Levites et al., 2000)

| Plant number | Phenotypic classes | | | |
| --- | --- | --- | --- | --- |
| | FF | FS | SS | 00 |
| Control | | | | |
| MsKW-35 | 105 | 107 | - | 13 |
| MsKW-11 | - | 21 | 2 | 7 |
| MsKW-13 | - | 79 | 8 | 18 |
| MsKW-30 | 26 | 44 | 13 | 2 |
| MsKW-33 | 15 | 20 | 3 | 2 |
| Colchicine treated | | | | |
| MsKW-88c | 25 | 26 | - | 4 |
| MsKW-11c | - | 101 | 23 | 28 |
| MsKW-1c | 69 | 115 | 23 | 16 |
| MsKW-9c | 34 | 49 | 10 | 17 |

The explanation of the observed changes should give an answer to the question of what a gene modified state is and the way this exchange proceeds. Searching an answer to the first question, it is necessary to note that practically all the plants classified as phenotypically modified do not carry an allele whose activity would be equal to null. Moreover, these plants are homozygous on the allele that preserves its expression. The supposition on the thing that nucleotide substitution could occur in the gene structure leading to the change of aminoacid and, hence, to the similarity of two initially different allelic enzyme variants, is rejected due to a very low probability of this event. Furthermore, the occurrence of similarity of two initially different ADH1 isozymes in agamospermous progenies is even less probable, as this gene alleles are different in two nucleotides, which are 27 nucleotides away from each other (Vinichenko et al., 2004). Suppose agamospermy leads to certain substitutions in locus *Adh1*. Then it is necessary to suggest that these alterations involve this gene region including not less than 29 nucleotide pairs. Then there is one more question: why only part of the gene is to change? Is the whole gene substitution possible?

Earlier we hypothesized that ovule cells with certain levels of polyteny tend to embryogenesis by mitotic agamospermy (Levites et al., 2000). Polyteny has been extensively reported of in many plant species (Kirianov et al., 1974; Nagl, 1976a, 1976b, 1981; Carvalheira, 2000; Guerra, 2001). Polyteny is frequently observed in generative organs, particularly in ovular and anther tapetum cells. Comparison of a rather well studied process of chromosome polytenization in *Diptera* and plants leads to the following observations. In *Diptera*, the somatic pairing of homologous chromosomes gives the false impression that there has been a decrease in chromosome number, because each nucleus appears to contain the haploid number of giant chromosomes (Bridges, 1935; Ashburner, 1970). In plants, endoreduplication leads to chromatid pairing, but this is not so close as in *Diptera*, and cell nuclei have a diploid number of polytene chromosomes. Furthermore, in the tapetum tissue, cells with nucleus having a polyploid number of chromosomes may occur, and so may those with nucleus having a diploid number of polytene chromosomes (Carvalheira, 2000).

It is possible that strict logic behind some types of ratios of phenotypic classes is conditioned by combinatorial processes that result in a few out of a total set of possible variants.

Based on the above reasoning, a model to explain the obtained results is proposed. Progeny msKW-30 (Table 3) was produced by agamospermy from a diploid heterozygous plant having genotype *Adh1-F/Adh1-S* (brief *FS*). This plant may have *FFFFSSSS* somatic cells in ovule tissues. This formula implies that there were eight chromatids, whose pairwise combination determines the genotype relative to locus *Adh1* of the cell entering embryogenesis and, consequently, genotype of agamospermy-produced embryo. This cell entering embryogenesis may be designated as initial proembryonal cell (IPC) or apozygote (APZ). Pairwise equally possible combination of eight elements shows that there are three genotypic classes possible at the ratio of 3*FF*:8*FS*:3*SS*. The experimentally revealed ratio of phenotypes in progeny msKW-30 is in good agreement with the expected value ($\chi^2$ =5.33). Pairwise combination of two chromatids does not imply that they reach the same pole. In this case, it only implies that these two chromatids stick to a factor like the nuclear membrane or a nuclear matrix. The conclusion about attachment of eukaryotic chromosomes to nuclear membrane was first made at first by A.N. Mosolow (1972). Now, this is a well known fact (Taddei et al., 2004). It is the attachment of chromatids to the nuclear membrane or a nuclear matrix that determines the genotype of initial proembryonal cell (IPC) or apozygote (APZ). The other six chromatids did not stick to that important factor and were eventually lost as the cell was undergoing division. Designating the chromatids sticking to the nuclear membrane or matrix as *F* or *S*, the processes can be schematically portrayed as follows:

*FFFFSSSS*→ *FF* (genotype of IPC), 2*F* and 4*S* to be lost soon

*FFFFSSSS*→ *FS* (genotype of IPC), 3*F* and 3*S* to be lost soon

*FFFFSSSS*→ *SS* (genotype of IPC), 4*F* and 2*S* to be lost soon

It is necessary to note that phenotypic ratio 3:8:3 can also be observed in meiotic agamospermy, previously designated as meiotic diplospory. According to the hypothesis of Maletskii and Maletskaya (1996), plant tissue mixoploidy underlies gametophytic agamospermy, i.e. the presence of tetraploid cell admixtures among the bulk of diploid cells. The existence of mixoploidy was demonstrated in many plants, especially in the *Chenopodiaceae* family including genus *Beta* (Gentcheff and Gustafsson, 1939; D'Amato, 1985; Carvalheira, 2000). Reductional division of admixed tetraploid cells results in the formation of diploid embryo sac with cells capable of embryogenesis. The 3:8:3 ratio of phenotypic classes observed in agamospermous offsprings is not different, in this case, from the one in gametes produced after meiosis under chromatid segregation in a tetraploid plant. Thus, a diploid seed progeny produced by meiotic agamospermy corresponds to the tetraploid plant gametes growing up to the mature seed state.

The way of determining the differences between meiotic and mitotic agamospermy at the phenotypic ratio 3:8:3 will be presented below.

Again, meiosis is in no way involved in the development of the studied progeny. Monomorphism of progeny msKWS1-5A relative to the heterozygous pattern of ADH1 (Table 1) provides further support to this statement. It is most likely that the stimulus to enter embryogenesis was an increase in chromosome or chromatid number by endopolyploidy or endoreduplication — similarly as in the case of gametes pairing. A combinatorial process determines the genotype of an IPC and losing an excess of the genetic material at the initial stages of embryo development.

This hypothesis is supported by the data obtained from analyzing barley zygotes (*Hordeum disticum* cv. Hauchen) at the initial stages of embryo development. DNA content in zygous nuclei of *H. disticum* was up to 16C, but it was gradually decreasing in proembryo nuclei down to 2C (Mericle and Mericle, 1970). The thing that a loss of genetic material during the first divisions of embryogenesis can occur was also well demonstrated in *Cyclopoida* and *Ciliatae* (Ammermann, 1971; Ammermann et al., 1974; Rasch and Wyngaard, 1997; Grishanin and Akif'ev, 1999). In our experiment, the evidence for this loss may be a large number of null phenotypes for ADH1 in agamospermous sugarbeet progenies (Table 3).

It is possible that a loss of ADH1 activity in some part of agamospermous progeny results from the loss of genetic material of *Adh1* alleles in all chromatids. It is noteworthy that the phenotype ratio of 3FF:8FS:3SS in progenies produced by mitotic agamospermy is quite rare. This could occur because the chromosome regions carrying different alleles of locus *Adh1* may be endoreduplicated frequently to different degrees. The possibility of unequal homological chromosome reduplication was shown earlier in beans (Cionini et al., 1982). A possibility of such asymmetry in plants we studied is supported by the fact that some seed progenies produced by agamospermy have only two phenotypic classes. An explanation could be found only if it was assumed that all the maternal plant somatic cells entering embryogenesis and contributing to embryos and seeds reproduction had the same genetic status, i.e. only one of the two alleles was endoreduplicated. For example, it is possible that msKW35 is the progeny of a heterozygous diploid plant having *FFFFFFSS* endoreduplicated cells in ovule tissues and msKW13 is the progeny of a heterozygous diploid plant having *FFSSSSSS* endoreduplicated cells, respectively. Occurrence of such a genotype can be considered as a consequence of the thing that only one pair of chromatids is reduplicated in each duplication cycle. The reason for this could be spatial limit in which only two chromatids can contact with the nuclear membrane or nuclear matrix out of the whole chromatid set which, due to such a contact, can reduplicate.

These processes of arising such genotype cells can be schematically portrayed as follows:

<u>FS</u> → <u>FF</u> SS → <u>FF</u> FF SS → <u>FF</u> FF FF SS → FF FF FF SS

Designating the paired *FF* and *SS* chromatids (chromosomes) as *F* or *S* and keeping in mind the possible role of paired *FF* and *SS* chromatids (chromosomes) attachment to the nuclear membrane, the processes can be schematically portrayed as follows:

***FFFS*** → <u>***FF***</u> (genotype of IPC), ***F*** and ***S*** to be lost soon

***FFFS*** → <u>***FS***</u> (genotype of IPC), 2***F*** to be lost soon

Thus, the agamospermous progeny of this heterozygous plant having ***FFFS*** endoreduplicated cells will consist of two genotypic classes *FF* and *FS* at the ratio 1:1 under equal probability of these two processes.

Likewise the genotypes of the embryos and seeds produced by agamospermy from a plant, in which endoreduplication had given rise to **FSSS** cells are determined.

**FSSS**→ **SS** (genotype of IPC), **F** and **S** to be lost soon

**FSSS**→ **FS** (genotype of IPC), 2**S** to be lost soon

Herewith, it is noteworthy once more that each randomly arising combination consists of two chromosomes, each having two chromatids. Hence, one can hypothesise that determination of initial proembryonal cell genotype occurs in this case due to the interaction with the nuclear membrane/matrix of already not two but four chromatids. It can be seen that the proposed model is based on the big role of chromatid and chromosome contact with the nuclear membrane.

This contact not only determines the character of phenotypic ratio in the progeny. Competition among chromatids (chromosomes) in the contact with nuclear membrane can be the source of different variability types in the seed progeny such as redetermination (Levites et al., 2001a).

There are more other arguments to support the thing that a combinatorial process in agamospermous way of reproduction involves a big chromatid number, and also there is a different degree of polyteny in chromosome regions carrying marker-genes. Thus, three phenotypic classes can be revealed in the same progeny on one marker enzyme and two on the other marker enzyme (Levites et al., 1999a). Moreover, in the agamospermous progeny of the hybrid ms SOAN41 x red table beet the phenotypic ratios of plant pollen sterility–fertility and colour well agree to the chromatid segregation model typical of meiotic agamospermy (Levites and Maletsky, 1999; Levites et al., 1999b). Further on, the genetic analysis of these plant genotypes on enzyme loci was carried out in the progeny obtained from analysing crosses. By means of malic enzyme pattern analysis, it was revealed that the produced agamospermous plants are represented only by two phenotypic classes, which is typical of not meiotic but mitotic agamospermy (Table 4). This ratio is significantly different from that of 3:8:3 typical of meiotic agamospermy (G = 225.4). For P=0.001, the critical value of G is 13.8.

Table 4. Malic enzyme phenotype ratio in agamospermous progeny of hybrid plant ms SOAN-41 x red table beet

| Analized form | Malic enzyme phenotype | | | G-test for 3:8:3 |
|---|---|---|---|---|
| | FF | FS | SS | |
| SOAN-41 x red table beet | 0 | 8 | 22 | 225.4 |

It may be considered as a genetic proof of the thing that cells entering embryogenesis have endoreduplicated chromosomes and there is different endoreduplication of chromosome regions carrying trait marker-genes.

The ratio of phenotypic classes in a progeny is equal both in mitotic and meiotic agamospermy under equal homological chromosome endoreduplication. However, if a phenotypic ratio corresponding to mitotic agamospermy is observed at least on one marker trait in the progeny, then all other phenotypic ratios in the same progeny should be considered as those corresponding to mitotic agamospermy. The presence of two phenotypic classes in a progeny is a more vivid evidence of the mitotic agamospermy.

The very existence of progenies that were produced by agamospermy and are represented by two phenotypic classes is interesting at least for it points out the specific underlying mechanism which controls this process.

Based on the concept that polymorphism in a progeny produced by mitotic agamospermy is due to chromosome endoreduplication, no polymorphism can be revealed unless endoreduplication is underway. That was what we observed in progeny msKWS1-5A produced by agamospermy: only plants heterozygous for *Adh1* were revealed. Thus, in this case the progeny resulted from a plant, in which both chromosomal regions carrying alleles *Adh1-F* and *Adh1-S* were not endoreduplicated. Nevertheless, the same progeny reveals dimorphism for IDH3 controlled by locus *Idh3* linked with the *Adh1* which is 17 cM away. This may be explained that agamospermous msKWS1-5A progeny comes from a heterozygous plant, in which the chromosome region with only one of the *Idh3* alleles is endoreduplicated. Hence, there is another important conclusion: different regions of a single chromosome can be endoreduplicated to different degrees. It well agrees to the data obtained in *Phaseolus cocineus* by morphological, autoradiographical and cytophotometric analyses (Cionini et al., 1982).

Differences in the endoreduplication degree of the same chromosome are largely conditioned by the thing that eukaryotic chromosomes contain many independent reduplication regions (Van't Hof, 1985, 1988; Bryant et al., 2001). Existence of underreduplicated and amplified (high polyteny level) chromosome regions was shown in many animal and plant species (Nagl, 1976a). However, the data obtained from *Phaseolus cocineus* (Cionini et al., 1982) and also the ratios of phenotypic classes for sugarbeet agamospermous progenies presented here allow us to speak of yet one more way of genetic encoding based on endoreduplication.

In the previous papers, it was proposed to regard genetic encoding based on endoreduplication as 2D (second dimension) of genetic encoding, whereas the genetic code written down as a nucleotide sequence - 1D (Levites, 2003, 2005). It was proposed also in these papers that the specific location of chromosomes in the cell nucleus be regarded as 3D encoding, and also there is a temporal dimension in encoding of inherited information in plants.

The role of spatial intra-nuclear chromosome location is well-known (Qumsiyeh, 1999; Leitch, 2000). Encoding inheritance in the 3D and temporal dimensions is normally perceived just as genome structural-functional organization including the dynamics of chromosome arrangement in the cell nucleus, chromatid state, time of DNA replication and gene expression (Judd, 1998; Leitch, 2000; Gribnau et al., 2003; Taddei et al., 2004; Donaldson, 2005).

It is easy to see that the genetic codes in all dimensions are differently affected by external and internal factors. Any nucleotide sequence could mainly change due to mutations, which happen rather rarely and are essentially nucleotide substitutions. Dependence of phenotypic ratios in an agamospermy-produced progeny on colchicine treatment, or on contribution by the maternal plant parents (Levites et al., 2000; 2001b) suggests that 2D genetic encoding depends on external and internal conditions. 3D encoding depends on such factors too and mostly on time. It is quite obvious that temporal encoding and its realization largely depend on external factors.

The assertion about involvement of this or that mechanism in genetic encoding requires a proof that it really functions during gamete formation.

There are direct proofs for endoreduplication in female gametes (egg cells). It is known that the DNA content of *Pinus sibirica* Du Taur egg cell nucleus is 16 times bigger than that of a somatic diploid cell and is equal to 32C (Ermakov et al., 1981), in *Ornithogalum caudatum* and *Haemantus albiflos*, the DNA content in egg cell nucleus is 4C and 3-4C, respectively (Morozova, 2002). The possibility of endoreduplication during plant gamete formation was also demonstrated in the research with *Allium tuberosum* (Kojima and Nagato, 1992). Herein it was shown that endoreduplicational meiosis is observed in females with the frequency of 80% and far less – 3.9% in males. In this case autobivalents formed by identical chromosomes occurring as a result of excessive pre-meiotic reduplication are observed. High DNA content in barley and petunia zygotes, gradually decreasing during the first zygote division and then reaching a diploid level at later stages of embryogenesis (Mericle, and Mericle, 1970; Vallade et al., 1978), is an indirect proof of chromosome endoreduplication in the egg cell nucleus.

Indirect data obtained from other embryo sac cells are also indicative of the thing that chromosome endoreduplication is possible in egg cells of many plant species. Thus, polytene chromosomes were revealed in synergieds of bear's onion (*Allium ursinum* L.) (Hasitschka-Jenschke, 1957) and nodding onion (*Allium nutans* L.) (Stozharova and Poddubnaya-Arnol'di, 1977). High DNA contetnt in the synergied nucleus of *Ornithogalum caudatum* and *Haemanthus albiflos* is also indicative of endoreduplication processes (Morozova, 2002). The presence of polytene chromosomes was shown in the antipodes of *Scilla bifolia* L. (Nagle, 1976), *Aconitum napellus* L., *Papaver heldreichii* Boiss, *Hypecoum procumbens* S.G. Gmel. (D'Amato, 1984). Moreovere, endoreduplication in embryosac cells (synergieds, antipodes) was shown in many plant species such as *Aconitum* (Tschermak-Woess, 1956), *Helleborus niger* (Hasitschka-Jenschke, 1959), *Ranunculus peltatus* Schrank. and *R. penicillatus* (Dumorf.) Bab. (Turala-Szybowska, 1980), *Ranunculus baudotii* Godr. (Wedzony, 1982), *Aquilegia vulgaris* L. (Turala-Szybowska and Wolanska, 1989), *Consolida regalis* L. (Unal and Vardar, 2006). A high degree of similarity for a number of traits and properties of synergieds, antipodal and egg cells, e.g. these cells capability of entering embryogenesis (Czapik, 1999; Batygina et al., 2003), are indicative of the thing that endoreduplication is possible in female gametes of many plant species.

There are more proofs of the thing that DNA synthesis proceeds in certain cell nuclei which do not undergo further division. It was shown in measuring the DNA content of both a vegetative cell nucleus and spermia cell nucleus in tobacco and tomato pollen grains (D'Amato et al., 1965). Similar results were obtained also in ten angiospermous species (Morozova et al., 1981; Morozova and Ermakov, 1993).

It well agrees to the data obtained in animals. Thus, for example, the DNA quantity of *Mesocyclops edax* spermatozoids is twice bigger than it could be expected based on DNA content in somatic cells of a diploid individual (Rasch and Wyngaard, 2001). The indirect proof of the thing that additional DNA synthesis can proceed in female gametes was obtained in insects. For example, there is additional chromosome duplication in female meiotic prophase in *Sipyloidea sipylus* (Pijnacker and Ferwerda, 1978).

On the basis of the above-mentioned facts of chromosome endoreduplication in sexual eukaryotic cells, one can hypothesise that the combinatorial process also proceeds in the genotype determination of progeny set under regular sexual reproduction. The earlier data we obtained on the absence of either maternal or paternal marker gene allele expression in a gamospermous progeny are the evidence of this. These facts were revealed when analyzing the seed progeny of plant (T1) taken from cv. Mezhotnenskaya 070 and used both as a maternal parent in its free pollination and as a pollinator in cross combinations: KHBC2-3A-24(*FC*) x T-1(*FF*), KHBC2-3A-36(*FC*) x T-1(*FF*), KHBC2-

3A-37(*FC*) x T-1(*FF*) (Table 5). Marker gene Me 1 having alleles F, S and C, each being present in cv. Mezhotnenskaya 070, was used in these experiments. A big set of phenotypic classes was revealed in the seeds set in plant T-1 under free intracultivar pollination(Table 5, Fig.2). The phenotypic ratio of these classes points out the thing that plant T-1 had the *Me-1F/Me1-F* genotype. However, the presence of phenotype CC in the progeny gathered from plant T-1 indicates the absence of maternal allele *Me1-F* expression in these seeds. This could be explained with the fact that allele *Me1-C* was introduced in the zygote by more than one chromatid. Besides, one should suggest that maternal allele *Me1-F* was also introduced in the zygote by more than one chromatid. Herein, pairwise chromatid combination is to result in the arising of genotypes *FF*, *FC* and *CC*.

Table 5. Malic enzyme phenotypes in hybrid progenies of sugarbeet plant T-1, genotype *Me1-F/Me1-F* (*FF*) (Levites, 2002b; Levites and Kirikovich, 2003)

| Analyzed form | Malic enzyme phenotypes | | | | |
|---|---|---|---|---|---|
| | CC | FC | FF | FS | CS |
| ♀T-1(*FF*) (free pollination) | 4 | 5 | 15 | 14 | 1 |
| ♀KHBC2-3A-24(*FC*) x ♂T-1(*FF*) | 1 | 12 | 17 | - | - |
| ♀KHBC2-3A-36(*FC*) x ♂T-1(*FF*) | 2 | 22 | 17 | - | - |
| ♀KHBC2-3A-37(*FC*) x ♂T-1(*FF*) | 2 | 16 | 15 | - | - |

The proposed mechanism outlines the aberration way of Mendel's law of uniformity in $F_1$ hybrids produced by crossing two constant homozygous forms. One can presuppose that expression absence of paternal allele introduced by T-1 pollen can, in many cases, be explained not by its silencing, but by the combinatorial process (Table 5).

Thus, the proposed model for 2D (second dimension) genetic encoding is based on the following statements:

1) Presence of unequal endoreduplication of chromosome regions within the same chromosome and absence of tight attachment of endoreduplicated chromatid regions.
2) Existence of the combinatorial process consisting in pairwise combination of homological chromatid regions.

3) Contact with the nuclear membrane or nuclear matrix of the developed pair of homological chromatid regions, due to which this pair is preserved in the following divisions of the cell (zygote or apozygote) entering embryogenesis. This pair of homological chromatids regions determines the embryo genotype.

The fact that endoreduplication frequency is different in female and male gametes (Kojima and Nagato, 1992) is undoubtedly of interest in estimating the role of endoreduplication in the processes of inheritance and variability. Such differences can be directly related to the expression of parental imprinting. At least the fact that plant imprinting was first investigated in maize endosperm cells having a specific maternal – paternal genomic ratio is indicative of this (Kermicle, 1978; Lin, 1982). Moreover, introduced by Johnston and coauthors (1980), the concept of Endosperm Balance Number which reflects the differences in the effective ploidy in the endosperm of studied crossed plant species may be explained now on the basis of differences of endoredupication level of chromosomes proposed here. In our experiments, imprinting expressed in the thing that the frequency of enzyme locus alleles variability in an agamospermous sugarbeet progeny depends on which proancestor this allele was introduced from in genome of maternal plant capable of agamospermous reproduction (Levites et al., 2001b).

Regarding the suggested model, it is interesting to consider well-known facts of plant epigenetic variability revealed in the researches of B. McClintock (1951, 1984), R.A. Brink (1960), E.H. Coe, Jr. (1966). It is possible to hypothesise that the genomic shock in wide hybridization that B. McClintock (1984) wrote about - accompanied by chromosome rearrangements and activation of mobile genetic elements - is the result of mutual adaptation of different genomes to the spatial arrangement within one cell nucleus. The model presented here allows us to suggest that the necessity for mutual adaptation is conditioned by a different degree of endoreduplication in chromosomes belonging to different genomes. In this case, activation of mobile elements can be considered as a compensatory process due to which moving of minor elements enables to preserve the arrangement and functional activity of major genome regions.

It is not excluded that the arising of big inverted repeats might be the result of detachment that begins on the ends of endoreduplicated region and the following chromatid arrangement in one line.

It is possible to hypothesise that many changes considered now as epigenetic are a consequence of specific genetic processes representing change genetic information in the 2D and 3D dimensions. The effect of such a wide-known mechanism as DNA methylation can be considered more likely as a consequence of the changes in 2D and 3D dimensions. This approach demonstrates that the inheritance of any seemingly monogenic trait is governed by many factors, which provide evidence that a clear-cut line between the discrete and continuous inheritance is absent.

This view of the encoding of genetic information may help overcome obstacles in the way towards understanding the mechanisms underlying inheritance of acquired traits. It is interesting to study a dependence of chromosome endoreduplication level on those external factors as quantity and quality of nutrition. It was shown earlier, the specific componential ratio of mineral nutrition in flax (Durrent, 1962; Durrent and Timmis, 1973) or treatment of germinating seeds with nicotinic acid (Bogdanova. 2003) can induce changes of plant morphological traits that preserve then for many generations. It was demonstrated that these arising phenotypic changes are accompanied by those of cell DNA content. Thus, for instance, an increase of DNA content (Bogdanova, 1992) was observed in wheat plants having preserved their traits of gigantism for 57 generations. It is not excluded that this increase of DNA content is conditioned by chromosome endoreduplication.

Considering the effect of colchicine treatment on the phenotypic ratio in an agamospermous progeny and the dependence of these ratios on the origin of marker gene alleles, also a DNA increase in stable inherited changes and the fact that replication wholly depends on nutrition and, particularly, sugar consumption (Riou-Khamlichi et al., 2000), it is possible to conclude that differential chromosome endoreduplication can be considered as a way of writing down of inherited information about acquired traits.

### ACKNOWLEDGEMENTS
I would like to express my gratitude to our Plant Populational Genetics Laboratory staffers as co-authors of our research referred to in this paper and Alexander V. Zhuravlev for the English version of this article.

### REFERENCES
**Ammermann, D. (1971).** Morphology and development of the macronuclei of the ciliates *Stylonychia mytilus* and *Euplotes aediculatus*. Chromosoma, **33**:209-238.

**Ammermann, D., Steinbruek, G., Berger, L. von and Hennig, W. (1974).** The development of the macronucleus in the ciliated protozoan *Stylonichia mytilus*. Chromosoma, **45**:401-429.

**Ashburner, M. (1970).** Function and structure of polytene chromosomes during insect development. Adv. Insect. Physiol., **1**:1-95.

**Asker, S.E. and Jerling, L. (1992).** Apomixis in plants. CRC Press, Boca Raton. 298 pp.

**Batygina, T.B., Bragina, E.A. and Vasilyeva, V.E. (2003).** The reproductive system and germination in Orchids. Acta Biol. Cracovensia Series Botanica, **45(2)**:21–34.

**Bogdanova, E.D. (1992).** Wheat genetic variability induced by nicotinic acid and its derivatives. Thesis of Dr. Sci. (Biol.). Novosibirsk: Institute of Cytology and Genetics Siberian Branch of Russian Academy of Sciences, 331pp.

**Bogdanova, E.D. (2003).** Epigenetic variation induced in *Triticum aestivum* L. by nicotinic acid. Russian Journal of Genetic, **39:**1029-1034.

**Bridges, C.B. (1935).** Salivary chromosome maps: with a key to the banding of the chromosomes of *Drosophila melanogaster*. J. Hered., **26**:60-64.

**Brink, R.A. (1960).** Paramutation and chromosome organization. Q. Rev. Biol., **35**:120-137.

**Bryant, J.A., Moore, K. and Aves, S.J. (2001).** Origins and complexes: the initiation of DNA replication. Journ. of Exp. Botany, **52**(355):193-202.

**Carvalheira, G. (2000).** Plant polytene chromosomes. Genet. Mol. Biol., **23(4)**:1043-1050.

**Czapik, R. (1999).** Enigma of apogamety. Protoplasma, **208**: 206-210.

**Cionini, P.G., Cavallini, A., Corsi, R. and Fogli, M. (1982).** Comparison of homologous polytene chromosome in *Phaseolus cocineus* embryo suspensor cells: morphological, autoradiographic and -cytophotometric analyses. Chromosoma, **86**:383-396.

**Coe, E.H., Jr. (1966).** The properties, origin and mechanism of conversion-type inheritance at the *B* locus in maize. Genetics, **53**:1035-1063.

**D'Amato, F. (1984).** Role of polyploidy in reproductive organs and tissues. In: B. M. Johri [ed.]. Embryology of angiosperms, P.518–566. Berlin, Heidelberg, New York: Springer-Verlag.

**D'Amato, F. (1985).** Cytogenetics of plant cell and tissue cultures and their regenerantes, CRC Crit. Rev. Plant Sci., **3(1)**:73-112.

**D'Amato, F., Derveux, M. and Scaracia-Mugnozza, D.T. (1965).** The DNA content of nuclei of pollen grains in tobacco and barley. Caryologia, **18(12)**:377-382.

**Donaldson, A. (2005).** Shaping time: chromatin structure and DNA replication programme. Trends in Genetics, **21(8)**:444-449.

**Dresselhaus, T. (2000).** Mendel, apomixis and plant breeding. Vortr. Pflanzenzuchtung, **48**:207-216.

**Durrent, A. (1962).** The environmental induction of heritable changes in Linum. Heredity, **17(1)**:27-61.

**Durrent, A. and Timmis, J.N. (1973).** Genetic control of environmentally induced changes in Linum. Heredity, **30(3)**:369-379.

**Ermakov, I.P., Barantseva, L.M. and Matveeva, N.P. (1981).** Cytochemical investigation of DNA during ovule development and early embryogenesis in *Pinus sibirica* Du Tour, Ontogenes, **12(4)**:339-345.

**Gentcheff, G. and Gustafsson, A. (1939).** The double chromosome reproduction in spinaceae and its causes: I. Normal behaviour. Hereditas (Lund, Swed.), **25(3)**:349-358.

**Gribnau J., Hochedlinger, K., Hata, K., Li E., and Jaenisch, R. (2003).** Asynchronous replication timing of imprinted loci is independent of DNA methylation, but consistent with differential subnuclear localization. Genes & Development, **17**:759-773.

**Grishanin, A.K. and Akif´ev, A.P. (1999).** Interpopulation differentiation within *C.kolensis* and *C.strenuus strenuus* (Crustacea: Copepoda): evidence from cytogenetic methods. Hydrobiologia, **417**:37-42.

**Guerra, M. (2001).** Fluorescent *in situ* hybridization in plant polytene chromosomes. Methods in Cell Science, **23**:133-138.

**Gustafsson, A. (1946-1947).** Apomixis in higher plants. Lunds. Univ. Arsskz. N.S. Sect.2 **42(3)**:1-67; **43(2)**:71-179; **43(12)**:184-370.

**Hasitchka-Jenschke, G. (1957).** Die Entwicklung der Sammenanlage von *Allium ursinum* mit besonderen Berucksichtigung der endopolyploiden Kerne in Synergiden und Antipoden, Oesterr. Bot. Z., **104(1/2)**:1-24.

**Hasitschka-Jenschke, G. (1959).** Vergleichende karyologische Untersuchungen an Antipoden. Chromosoma **10**:229-267

**Johnston, S.A., den Nijs, T.P.M., Peloquin, S.J. and Hannenman, R.E., Jr. (1980).** The significance of genic balance to endosperm development in interspecific crosses. Theor. Appl. Genet., **57**:5-9.

**Judd, B.H. (1998).** Genes and chromomeres: a puzzle in three dimensions. Genetics, **150**:1-9.

**Kermicle, J.L. (1978).** Imrinting of gene action in maize endosperm. In: D.B.Walden [ed.]. Maize Breeding and Genetics, P.357-371. N.Y.: John Wiley & Sons.

**Khohlov, S.S. (1967).** Apomixis: classification and diffusion in angiospermous plants. Advances in Modern Genetics, **1**:43-105 (In Russian).

**Kirianov, G.I., Polyakov, V.Yu. and Chentsov, Yu.S. (1974).** Biochemical approach to the problem of some plant chromosome polynemity. DokladyAkad. Nauk. **218(2)**:485-488 (In Russian).

**Kojima, A. and Nagato, Y. (1992).** Diplosporous embryo-sac formation and the degree of diplospory in *Allium tuberosum* Biomedical and Life Sciences, **5(1)**:72-78.

**Koltunow, A.M. (1993).** Apomixis: embryo sacs and embryos formed without meiosis or fertilization in ovules. Plant Cell, **5**:1425:1437.

**Leitch, A.R. (2000).** Higher levels of organization in the interphase nucleus of cycling and differentiated cells. Microbiol. Mol. Biol. Rev., **64(1)**:138-152.

**Levites, E.V. (1979).** Genetic control of NADP-dependent malic enzyme in sugarbeet (*Beta vulgaris* L.). Doklady Biol. Sciences, **249**:1241-1243.

**Levites, E.V. (2002a).** New classification of the reproduction modes in sugar beet. Sugar Tech, **4(1&2)**:45-51.

**Levites, E.V. (2002b).** Redetermination: an interesting epigenetic phenomenon associated with mitotic agamospermy in sugar beet. Sugar Tech, **4(3&4)**:137-141.

**Levites, E.V. (2003).** Theoretical and practical aspects of studies in epigenetic variability in sugarbeet. Sugar Tech, **5(4)**:209-211.

**Levites, E.V. (2005).** Sugarbeet plants produced by agamospermy as a model for studying genome structure and function in higher plants. Sugar Tech, **7(2&3)**:67-70.

**Levites, E.V., Denisova, F.Sh., Kirikovich, S.S. and Judanova, S.S. (Maletskaya, S.S.) (2000).** Ratios of phenotypes at the *Adh1* locus in the apozygotic offspring in sugarbeet ($C_1$ generation). Sugar Tech, **2(4)**:26-30.

**Levites, E.V., Denisova, F.Sh. and Filatov, G.P. (1994).** Genetic control of isozymes in the sugar beet. In : C.L.Markert, J.G.Scandalios, H.A.Lim and O.L.Serov [eds.]. Isozymes: organization and roles in evolution, genetics and physiology, P. 171–178. Singapore. New Jersey. London. Hong Kong: World scientific.

**Levites, E.V. and Kirikovich, S.S. (2003).** Epigenetic variability of unlinked enzyme genes in agamospermous progeny of sugarbeet. Sugar Tech, **5(1&2)**:57-59.

**Levites, E.V., Kirikovich, S.S. and Denisova, F.Sh. (2001b).** Expression of enzyme genes in agamospermous progenies of reciprocal hybrids of sugar beet. Sugar Tech, **3(4)**:160-165.

**Levites, E.V., Kudashova, T.Yu. and Viksler, L.N. (1988).** Study of syntene gene groups in sugarbeet. Novosibirsk:Institute of Cytology and Genetics of SB AS SSSR, 24 pp.

**Levites, E.V., and Maletskii, S.I. (1999).** Auto- and episegregation for reproductive traits

in agamospermous progenies of beet *Beta vulgaris* L. Russian Journal of Genetics, **35(7)**:802-810.

**Levites, E.V., Ovechkina, O.N. and Maletskii, S.I. (1999b).** Auto- and episegregation for color types in agamospermous progenies of beet (*Beta vulgaris* L.). Russian Journal of Genetics, **35(8)**:931-936.

**Levites, E.V., Shkutnik, T., Ovechkina, O.N. and Maletskii, S.I. (1999a).** Pseudosegregation in the agamospermic progeny of male sterile plants of the sugar beet (Beta vulgaris L.). Doklady Biol. Sciences, **365**:182-184.

**Levites, E.V., Shkutnik, T., Shavorskaya, O.A. and Denisova, F.Sh. (2001a).** Epigenetic variability in agamospermous progeny of sugar beet. Sugar Tech, **3(3)**:101-105.

**Lin, B.-Y. (1982).** Association of endosperm reduction with parental imprinting in maize. Genetics, **100**:475-486.

**Maletskii, S.I. (1999).** Epigenetical variability of the unianthy and synanthy expression in sugarbeet. Sugar Tech, **1(1&2)**:23-29.

**Maletskii, S.I. and Konovalov, A.A. (1985).** The inheritance of alcohol dehydrogenase in sugarbeet. I. The analysis of deviations from monogene segregation. Genetika, **21(9)**:1527-1534 (In Russian).

**Maletskii, S.I., Levites, E.V., Baturin, S.O. and Yudanova, S.S. (2004).** Reproductive biology of angiospermous plants. Genetical glossary. Novosibirsk: Institute of Cytology and Genetics, 106 pp.

**Maletskii, S.I. and Maletskaya, E.I. (1996).** Self-fertility and agamospermy in sugar beet. *Beta vulgaris* L. Russ. Jour. Genetics, **32(12)**:1643-1650.

**Maletskaya, E.I. and Maletskaya, S.S. (1999).** The nuclear DNA mass variability in embryo root cells of sugarbeet. Sugar Tech, **1(1&2)**:30-36.

**Matzke, M.A., Mittelsten Scheid, O. and Matzke, A.J.M. (1999).** Rapid structural and epigenetic changes in polyploid and aneuploid genomes. Bio Essays, **21**:761-767.

**McClintock, B. (1951).** Chromosome organization and genic expression. Cold Spring Harbor Symp. Quant. Biol., **16**:13-47.

**McClintock, B. (1984).** The Significance of Responses of the genome to challenge. Science, **226**:792-801.

**Meizel, S. and Markert, C.L. (1967).** Malate dehydrogenase isozymes of the marine snail *Ilyanassa obsolete.* Arch. Biochem. Biophys., **122**:753-765.

**Mericle, L.W. and Mericle, R.P. (1970).** Nuclear DNA complement in young proembryos of barlei. Mutat. Res., **10(10)**:508-518.

**Morozova, E.M. (2002).** Additional nuclear DNA in cells of embryo sacs of *Haemanthus albiflos* and *Ornithogalum caudatum.* Biol. Bulletin, **29(2)**:192-195.

**Morozova, E.M. and Ermakov, I.P. (1993).** Cell cycle during development of male and female gametophytes in angiosperms. Physiol. and Bioch. of Cultivated Plants, **25(3)**:297-302 (In Russian).

**Morozova, E.M., Ermakov, I.P. and Karpova, L.M. (1981).** Cytophotometric studies of nuclei of mature pollen grains in some angiosperms. Biol. Nauki, **11**:71-76 (In Russian).

**Mosolow, A.N. (1972).** New approach to the solving of the problem of chromosomes space arrangement in the interphase nucleus (polar model of interphase nucleus). Tsitologia, **14(5)**:542-552 (In Russian).

**Nagl, W. (1976a).** Nuclear organization. Ann. Rev. Plant Physiol, **27**:39-69.

**Nagl, W. (1976b).** The polytenic antipodal cells of *Scilla bifolia*; DNA replication pattern and possibility of nuclear DNA amplification. Cytobiol., **14(1)**:165-170.

**Nagl, W. (1981).** Polytene chromosomes of plants. Int. Rev. Cytol., **73**:21-53.

**Pijnacker, L.P. and Ferwerda, M.A. (1978).** Additional chromosome duplication in female meiotic prophase of *Sipyloidea sipylus* West wood (Insecta, Plasmida), and its absence in male meiosis. Experientia, **34**:1558-1560.

**Qumsiyeh, M.B. (1999).** Structure and function of the nucleus: anatomy and physiology of chromatin. CMLS Cell. Mol. Life Sci., **55**:1129-1140.

**Rasch, E.M. and Wyngaard, G.A. (1997).** Analysis of DNA levels during gonomery in early cleavage divisions of the freshwater copepod *Mesocyclops edax*. Microsc. Microanal., **3**:191-192.

**Rasch, E.M. and Wyngaard, G.A. (2001).** Evidence for endoreduplication: germ cell DNA levels prior to chromatin diminution in *Mesocyclops edax*. J. Histochem. Cytochem., **49**:795-796.

**Richards, A.J. (1986).** Plant breeding system. Chapter 11, P. 403-456. London: Allen and Unwinn.

**Riou-Khamlichi, C., Menges, M., Healy, J.M.S. and Murray, J.A.H. (2000).** Sugar control of the plant cell cycle: differential regulation of *Arabidopsis* D-type cyclin gene expression. Molec. Cell. Biol., **20(13)**:4513-4521.

**Stozharova, I.A. and Poddubnaya-Arnol'di, V.A. (1977).** Role of synergids in fertilization and embryo-endospermiogenesis in *Allium nutans* L. Byull. Glavn. Bot. Sada Akad. Nauk SSSR, **105**:70-78 (In Russian).

**Taddei, A., Hediger, F., Neumann F.R. and Gasser, S.M. (2004).** The function of nuclear architecture: a genetic approach. Annu. Rev. Genet., **38**:305-345.

**Tschermak-Woess, E. (1956).** Notizen uber die Riesenkerne und 'Riesenchromosomen' in den Antipoden von *Aconitum.* Chromosoma, **8**:114-134.

**Turala-Szybowska, K. (1980).** Endopolyploidy in the antipodals of *Ranunculus peltatus* Schrank. and *R.penicillatus* (Dumorf.) Bab. Acta Biol. Cracoviensia, **22**:163-173.

**Turala-Szybowska, K. and Wolanska, J. (1989).** Two mechanisms of polyploidization of the antipodals in *Aquilegia vulgaris* L. Acta Biol. Cracoviensia Series Botanica, **31**:63-74.

**Unal, M. and Vardar, F. (2006).** Embryological Analysis of *Consolida regalis* L. (Ranunculaceae). Acta Biol. Cracoviensia Series Botanica, **48(1)**:27-32.

**Vallade, J., Cornu, A., Essad, S. and Alabouverte, J. (1978).** Niveaux de DNA dans les noyaux zygoteniques chez Ie *Petunia hybrida* h-ort. Bull. Soc. Bot. France, **125(1/2)**:253-258.

**Van't Hof, J. (1985).** Control points within the cell cycle. In: J.A. Bryant and D. Francis [eds.]. The cell division cycle in plants, P. 1–13. Cambridge, UK: Cambridge University Press.

**Van't Hof, J. (1988).** Functional chromosomal structure: the replicon. In: J.A. Bryant and V.L. Dunham [eds.]. DNA replication in plants. P. 1-15. Boca Raton, FL, USA: CRC Press.

**Vinichenko, N.A., Golovina, K.A., Blinov, A.G., Sokolova (Antonova), O.O. and Levites, E.V. (2004).** Molecular differences between alleles *Adh1-F* and *Adh1-S* in sugarbeet *Beta vulgaris* L. Russian Journal of Genetics, **40(2)**:172-177.

**Wedzony, M. (1982).** Endopolyploidy and structure of nuclei in the antipodals and synergids of *Ranunculus baudotii* Godr. Acta Biol. Cracoviensia Series Botanica, **24**:43-62.